# Using Ontologies for the Design of Data Warehouses

Jesús Pardillo[1] and Jose-Norberto Mazón[2]

University of Alicante – DLSI/Lucentia, Spain
[1]`research@jesuspardillo.com`
[2]`jnmazon@dlsi.ua.es`

## ABSTRACT

*Obtaining an implementation of a data warehouse is a complex task that forces designers to acquire wide knowledge of the domain, thus requiring a high level of expertise and becoming it a prone-to-fail task. Based on our experience, we have detected a set of situations we have faced up with in real-world projects in which we believe that the use of ontologies will improve several aspects of the design of data warehouses. The aim of this article is to describe several shortcomings of current data warehouse design approaches and discuss the benefit of using ontologies to overcome them. This work is a starting point for discussing the convenience of using ontologies in data warehouse design.*

## KEYWORDS

*Data Warehouses, Ontologies*

## 1. INTRODUCTION

Data warehouse systems provide a multidimensional view of huge amounts of historical data from operational sources, thus supplying useful information, which allows decision makers to improve business processes in organizations. A multidimensional model structures information into facts and dimensions. A fact contains the interesting measures (fact attributes) of a business process (sales, deliveries, etc.), whereas a dimension represents the context for analyzing a fact (product, customer, time, etc.) using hierarchically organized dimension attributes. Multidimensional modeling requires specialized design techniques that resemble the traditional database design methods [1]. First, a conceptual design phase is carried out, whose output is an implementation-independent and expressive conceptual multidimensional model for the data warehouse. A logical design phase then aims to obtain a technology-dependent model from the previously defined conceptual multidimensional model. This logical model is the basis for the implementation of the data warehouse. Therefore, the multidimensional design is driven by models to reflects real-world scenarios and obtain the most suitable logical representation.

One benefit of multidimensional modeling has been to increase the level of automatization in obtaining a final implementation at the same time that semantic gap among different models are bridged. To this aim, so far, we have been applying UML and MDA in the design of data warehouses [2, 3, 4]. More concretely, we have developed a hybrid approach to reconcile the conceptual schemas obtained from user's requirements and the conceptual schemas from data sources [5]. We have also developed a prototype (http://www.lucentia.es) based on the Eclipse Modeling Framework that allows us to apply our approach in real-world projects.

However, obtaining an implementation of the data warehouse is a complex task that often forces designers to acquire wide knowledge of the domain, thus requiring a high level of expertise and

                                                                                                                 73

becoming it a prone-to-fail task. Based on our experience, we have detected a set of situations (*e.g.*, additivity, reconciling requirements and data sources, conformed dimensions, etc.) we have faced up with in real-world projects in which we believe that the use of some kind of knowledge resources will improve the design of data warehouses (some of them are being considered). In the light of these issues, ontologies seem to be a promising solution, since they are common conceptualization of a domain, representing shared knowledge: while models are prescriptive, ontologies are descriptive [6], which means that models state requirements about the system-to-be (*e.g.*, what and how should it be built), and ontologies describe the application domain as it is (in a technology-independent manner).

Importantly, ontologies may empower the automatization advocated by model-driven development, since they provide mechanisms to formally specify the semantics of a domain on which models may be supported. Interestingly, we argue that ontologies could be used in data warehouse development since data structures depend on a given context to define their actual semantics, and ontologies provide the crucial context knowledge relevant to interpret semantics [7]. For instance, the fact *sales* may be interpreted as (i) "a collection of actions and a specialization of business activity and commercial activity" (the conventional sense) or as (ii) "the collection of *OfferingForSale* events including events in which an agent offers one or more things for sale to one or more agents" (http://www.cycfoundation.org/concepts). Moreover, in the context of data warehousing, *sales* may refer to both punctual and cumulative sales (in the sense of snapshot stocks). Finally, if *sales* are modeled as a fact of a data warehouse and we find different sales' concepts in data sources with the same functional dependencies, we cannot automatically decide which one to be taken without considering semantic knowledge. In each case, its management and analysis imply different treatments.

Therefore, in this article, we present a set of situations in which ontologies may help data warehouse designers to solve several problems. On the same way, there are other authors who have also started the use of ontologies in different aspects of a data-warehouse architecture such as ETL processes [8] or data sources [9]. However, there is a still a plethora of situations where ontologies may be applied and there is not still a research roadmap on how to do it. This article sheds light on the benefit of using ontologies in data warehouse research.

This article describes several shortcomings of current data warehouse design approaches and how ontologies can help in overcoming them. This work is intended to be an avenue for future work as we pretend to give hints and start a discussion about the convenience of using ontologies in data warehouse design.

## 2. ONTOLOGY-AWARE DATA WAREHOUSE DESIGN

In the following, the situations in which ontologies may help designers in the development of data warehouses are presented as follow: (i) *Shortcoming:* the situation is presented as a shortcoming followed by a short definition of the detected problem, (ii) *Description:* a deeper description of the problem is presented to clarify the found shortcoming, and (iii) *Discussion:* a discussion on how ontologies may help designer in improving the current solution for the found problem is presented.

### 2.1. Shortcoming: Requirement Analysis in Multidimensional Design

*Requirement analysis for data warehouse design needs new concepts and techniques whose meaning should be clarified to be used by stakeholders.*

### 2.1.1 Description

Requirements analysis for multidimensional modeling pretends to elicit information needs of decision makers to deploy a data warehouse that satisfy their expectations. Data warehouse design needs for new concepts and techniques different from those used in traditional requirement engineering tasks. Currently, several proposals for eliciting information requirements in data warehouses have been developed [10, 11, 12]. However, these approaches ignore how to formally describe the meaning of every decision concept and how to share this knowledge to be easily used by requirement engineers.

### 2.1.2. Discussion

The role of a business model is highly important in requirement engineering to every stakeholder understands one another. This is especially useful because stakeholders may have incompatible needs [13]. Importantly, business models should be managed as an ontology, because they are descriptive abstraction of the environment in which any software system (including a data warehouse) should operate. They also contain the semantics already agreed by the organization stakeholders. Moreover, an ontology such as the one presented in [14] that is based on speech acts [15] may serve as a guide for analyzing information requirements in an ontological way. As its authors state "the only relevant core ontology is one which helps the software engineer in solving the requirements problem" [14].

Multidimensional concepts and techniques should be clarified when a requirement analysis stage is applied to the development of data warehouses. For instance, a foundational ontology such as the Bunge's one [16] serves us for validating multidimensional models [3, 17] by a *representation mapping* where ontological concepts are mapped into language constructs and an *interpretation mapping* where for the later ones an ontological interpretation is assigned.

Thus, ontologies are not the panacea for requirement engineering in data warehouses but they are an interesting complement for formalizing what is really meaning by stakeholders when multidimensional issues arise.

### 2.2. Shortcoming: Reconciling Requirements and Data Sources

> *Due to the special idiosyncrasy of data warehouses, not only requirements should be considered for multidimensional design, but also a second driving force: data sources, which need to be reconciled with information requirements.*

### 2.2.1. Description

Requirements analysis for multidimensional modeling pretends to elicit information needs of decision makers to deploy a data warehouse that satisfy their expectations. Furthermore, available data sources should be taken into account since they will populate the data warehouse. Currently, several proposals for eliciting information requirements in data warehouses have been developed [11, 12] to derive multidimensional models that describe real world in terms of facts and dimensions. However, mechanisms through which to formally match the data sources with information requirements in early stages of the development are not investigated so far, thus the correspondence between information requirements and their counterparts in data sources are not obvious. The identification of multidimensional elements in the data sources is a mandatory previous step before reconciling requirements and data sources (*i.e.*, fact, dimensions, bases, etc.).  These elements are usually annotated in a manual [10] or semiautomatic [18, 19] manner from data sources using syntactic information [20], which is not enough for every scenario and prevents their total automatization. In order to overcome this situation, we argue that the most promising solution consists of enriching the previous cited approaches by using the semantic knowledge provided by ontologies as stated in [9].

### 2.2.2. Discussion

One approach closely related to the idea of using ontologies for deriving multidimensional data is [9]. It presents a method for discovering multidimensional structure from ontologies. This method consists on: (i) discovering dimensions and measures by matching and subsumption, (ii) selecting measures and dimensions for defining facts, (iii) defining bases by exhaustive searching and pruning, and (iv) defining aggregation hierarchies by identifying part-whole relationships. Dimensional elements (*i.e.*, fact, dimensions, bases, etc.) are identified using heuristics based on structural aspects (such as counting instances, inserting frequency, cardinality, etc.). Other approach using heuristics is [18] where the fact-dimension dichotomy is based on data cardinality.

However, semantics about multidimensional properties of a given concept should also be considered. In this sense, the interesting debate comes from discovering the ontological foundation of multidimensional modeling. It may be articulated by answering if the proposed structural heuristics are pointing out semantics for which multidimensional elements will be univocally identified. Therefore, research community should work on providing both the involved semantics rules and the (public) repository of multidimensional-annotated ontologies. For the last one, an interesting starting point is the CyC ontology (http://www.cyc.com) or its open version OpenCyC (http://www.opencyc.org). CyC is an ontology conceived to answer common sense reasoning by computers. Let it be, both requirement-driven and data-driven methods could take advantage of the ontological knowledge about multidimensionality.

## 2.3. Shortcoming: Incompleteness in Multidimensional Models

> *Ontological knowledge may enrich a multidimensional model in aspects that have not been taken into account during requirement analysis or data-source alignment.*

### 2.3.1. Description

Data sources of a data warehouse may not cope with all information that end-user analysis requires. For instance: *product* dimension can be populated with data only for the product details themselves (*e.g.*, code, name, description). However, required additional aggregation levels such as product *subtype, type or class* may be not contained in operational data sources. Therefore, additional knowledge is required to deal with these new aggregation levels.

### 2.3.2. Discussion

There are some kinds of data that can be directly obtained from other, "public" data sources. For example, an ontology such as WordNet can be used to complete the unsupported elements within a dimension hierarchy. Moreover, taxonomies such as the *Computing Classification System* (CCS, http://portal.acm.org/ccs.cfm?part=author&coll=portal&dl=GUIDE) can be also used as additional data sources from which designing multidimensional models. CSS provides a classification for topics on computations that can be, (and usually is), used for classifying computer literature. For instance, there is a taxon for *Information Systems* (H) where, *e.g.*, *Database Management* (H.2) topics are classified. This taxonomy can also be used for completing aggregation hierarchies, whenever a data warehouse requires a dimension on computer-related topics. An example of using ontologies for enriching dimension hierarchies is defined in [21]: for the hierarchy city-state-country, the meronymy/holonymy relationship can be used due to the fact that city is a part of state and state is a part of country (*e.g.*, Boston is a part of Massachusetts, and Massachusetts is a part of USA), and if the hierarchy is product-family-class, hypernymy/hyponymy relationships can be used, because of every product is a kind of family and every family is a kind of class (*e.g.*, cake is a kind of baked good, and baked good is a kind of food).

## 2.4. Shortcoming: Data Types of Measures

*Measures of analysis described by multidimensional models in facts, even at the conceptual level, do not aid neither designers nor end-users to know important details such as their units or scale.*

### 2.4.1. Description

An interesting improvement in the semantic description of measures in multidimensional models is the addition of units, magnitudes, and scales. Regarding this issue, [22] proposed the well-known "levels of measurement": four different types of numerical scales, which he called *nominal*, *ordinal*, *interval* and *ratio*. This classification is interesting for multidimensional models because from these scale types, some statistics functions (and particularly, the aggregation functions used in OLAP systems) are restricted. Each scale type defines a mathematical structure on which the permissible statistics and scale transformations are allowed. For instance, nominal measures of a multidimensional model may be aggregated using *mode* or *chi-square* statistics and only directly compared using *equality*. However, in a more powerful scale such as interval, statistics such as *mean*, *standard deviation*, *correlation*, etc. are allowed.

### 2.4.2. Discussion

Some simple definition as units of measurement may dramatically empower the common definition of OLAP measures. Units of measurement are a common forgotten part of measurement definition. This topic directly deals with the ontological concept of *quantity* (http://www.wikipedia.org): *(it) is a kind of property, which exists as magnitude or multitude (number)*. Identifying these concepts in the multidimensional models, by their ontological alignment, enables designers to describe properly what is really modeled, and therefore, how actually should be analyzed.

Many measurement ontologies may aid designers to annotate the identified measures of OLAP analysis according to them. A well-known ontology in the software engineering field is the agreed by [23]. This ontology is defined by many representative research groups in the area of software measurement and quality and thus, provides commitment among them for uniformly specify measures. Thus, since data warehouses deal with measurement of facts for being analyzed, it is also useful in this context. In addition, ontologies for describing units of measurement domain may be also helpful: *e.g.*, the *Measurement Units Ontology* (MUO, http://idi.fundacionctic.org/muo/muo-vocab.html) or others from the biomedical domain such as the *units of measuremen*t (unit.obo, http://www.obofoundry.org/cgi-bin/detail.cgi?id=unit; as seen in Figure 1).

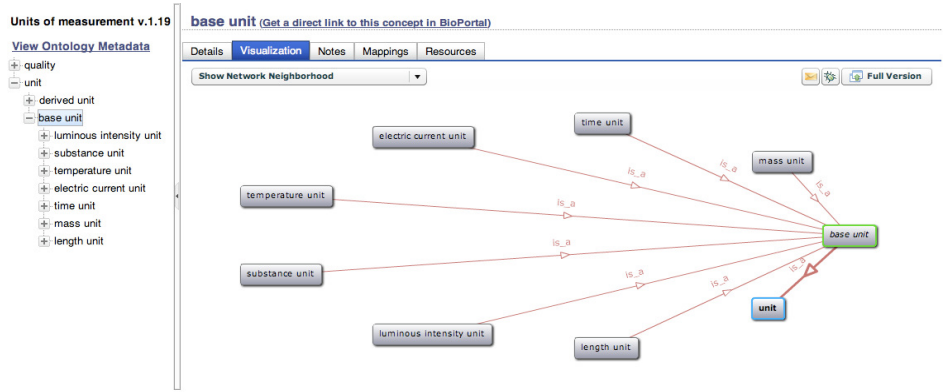

Figure 1. Visualization of an excerpt of the units of measurement ontology

### 2.5. Shortcoming: Semantic-aware Summarizability

*Many statistic functions may be used to aggregate data cells of measures. Their application depends on which sort of measure and aggregation criteria are involved.*

### 2.5.1. Description

Specifying additivity constraints is a fundamental part of a multidimensional model. In particular, type compatibility [24] states the compatibility of category attributes (aggregation levels), summary attributes (measures) and statistical functions (aggregation functions). A classification for measures is proposed in [25]: the well-known categories *additive*, *semi-additive* and *non-additive*. Despite of the fact that these categories are related to assuring additivity along all, some, or none dimensions, respectively, without taking into account their nature (semantics), other authors propose frameworks where measures are described by richer classifications. Take [24, 26] as example. In [24], authors propose the classification of summary attributes into can be classified as either a "flow" (also called "rate"), a "stock" (also called "level") or a "value-per-unit". In addition, [26] presents a taxonomy of summary constraints where, *e.g.*, non-additive measures are further classified into ratios and percentages, measures of intensity, or average/maximum/minimum together with some treatments for dealing with them. That is, there is not a common agreement.

### 2.5.2. Discussion

The proposed taxonomies for dealing with summarizability constraints lack in an ontological foundation. This means that designers do not have aids for reasoning about what *really* is a given piece of information and thus, which constraints should be hold. For instance, the concept expressed by "flow" in [24] and the concept expressed by "measure of intensity" in [26] may be the same.

The problem is that, without an ontological foundation (whether it be implicitly taken into account or explicitly codified in an ontology thought as a communication artifact) acting as a comparison framework, there is no possibility of being sure that these frameworks are complete or formally stated. Simply, there is only tacit knowledge, not formal semantics (two readers may interpret different things). Even more importantly, we cannot know why it is in that way and how sound our knowledge is.

In this way, ontologies enable us to gain better understanding about the very nature of summarizability constraints, reasoning about why, when, and how to hold it. One first issue to deal with is the role that time dimension plays in the summarizability constraints. Despite of the fact that type incompatibilities are not always associated to time dimension [17], time is a (probably, the) key dimension in a data warehouse. Indeed, data warehouses are *time-variant* [27] databases, *i.e.*, collection of time-stamped facts. Therefore, the first dimension to check additivity is usually this one. Here, these facts that are described temporally, may be understood as *events*, should be described by a foundational ontology such as DOLCE (http://www.loa-cnr.it/DOLCE.html) as *endurant* or *perdurant particulars*. The distinction between endurants and perdurants plays a prominent role in top-level ontologies and it can provide very useful knowledge about what is really managed by data warehouses and how OLAP analyses should be performed over them. These categories may be characterised by the following properties:

- Endurant: Particulars may be categorized as *endurants* iff (i) they persist, (ii) necessarily lack proper temporal parts, and (iii) are necessarily and completely present in each time interval at which they exist.

- Perdurant: Particulars may be categorized as *perdurant* iff (i) they persist, (ii) necessarily have proper temporal parts, and (iii) are necessarily not wholly present in each time interval at which they exist.

Given these (on the other hand, informal) definitions, a measure "sale units" may be interpreted as endurant sale, *i.e.*, the sale accomplished in a temporal moment, and as perdurant sale, *i.e.*, the (cumulative) sale that "lives" till now. Another example, the classical "inventory stock" (in the traditional sense) is not more, accordingly to the ontological basis, a perdurant entity. Therefore, by the semantic annotation of measures, its nature regarding time can be assessed. Since perdurant particulars live in a time period, they should be characterized as semi-additive measures along time dimensions of a data warehouse. Conversely, endurant particulars are attached to particular temporal moment, thus they can be added along time without ontologically-founded duplicated values.

Interestingly, ontologies allow us to establish endurant-perdurant mappings: because perdurant particular are attached to particular temporal points, two of them may be translated into an endurant one, and conversely, an endurant particular may be mapped into at least two perdurant particulars marking the beginning and end of the period over which exists. These facts provide real guides for checking and assuring summarizability constraints regarding time.

For instance, in order to make inventory stocks additive along time, we may map them into the input goods and output ones. Note that additivity, *i.e.*, the aggregation by *sum* function, is a desirable property of any measure because it is transitive along aggregates (data cubes) in contrast to other statistics such as means that should be recalculated from the most granular data. Then, we specify a derived measure, over which we will calculate by aggregating it from the previous ones by simply adding quantities. This measure can be calculated for obtaining the stock in given time period. Please note that conversely, "well-formed" sales (endurant particular) may be shown as "stocks being output". In this line, product returns should be "stocks being input" in the general balance of the company account. However, here, this conversion only implies harder constraints over summarizability, thus designers naturally overlook it. Interestingly, facts such as these can be logically derived from the ontologies. The method for gaining such results is the semantic annotation of multidimensional models. This is called "interpretation mapping" [28], that is the process of assigning ontological meaning to the modeling elements which have not yet one assigned. After this step, ontological reasoners may be used to infer facts to both aiding to define summarizability constraints and evaluate how sound are the constraints manually imposed.

### 2.6. Shortcoming: Conformed Dimensions

*Data-warehousing architecture deals with data marts to customize the access to the corporate data warehouse. However, in order to tie data marts together, they should be conformed.*

### 2.6.1. Description

The orchestration of the design of databases for the data warehouse repositories is recognized in terms of the architectural debate [29]. There are two different points of view: on the one hand, Inmon [27] advocates the design of corporate data warehouses that elegantly integrate all data sources which will be then mapped into departmental data marts. On the other, Kimball [25] advocates the design of data marts (related by a "bus architecture") that may be additionally supported by a data warehouse.

### 2.6.2. Discussion

Whichever approach is selected, there is a step in the development process where both structural and semantics mapping are involved. Despite of the fact that works such as [30, 31] deal with the integrated design of data marts and data warehouses, they are mainly oriented to the structural design of databases. With regard to the structural mappings, data marts models are summarized or customized versions of the data warehouse counterparts where aggregation hierarchies are pruned and dimensions are limited (acting as materialized views). However, the integration of such data structures also involves semantics mappings. A data mart should be integrated (or conversely, a data warehouse should be customized as a data mart) by matching modeling elements. Moreover, data marts should be tied together to be able to reply cross-departmental queries elaborated from OLAP operations such as the classical drill across [25]. This situation, *i.e.*, data mart and data warehouse integration, indeed arise problems related to semantics. Otherwise, integrating them is one hardest problem in real-world projects.

In particular, conformed dimensions mean the 80% of effort in the development of data marts [25]. Current approaches are based on some variant of the foundational definition of conformity provided by Kimball: "two dimensions are conformed if the fields that you use as common row headers have the same domain" [32]. These definitions manage conformity as similarity functions of the meaning, *i.e.*, ontological entities of multidimensional models. Thus, a promising solution involves ontologies for abstracting the semantic layer on which both data marts and corporate data warehouses (in the sense of data structures) are supported. For instance, a sample process would act as follows: the semantic annotation of each modeling element regarding a domain ontology is performed first. This step involves every repository in the architecture (data marts and data warehouse). Second, subsumption reasoning is used for inferring proper ways to match them. Third, given the mapping output from the (automatic) reasoner, a data transformation is used for implementing the transference among repositories, acting in the same sense of ETL processes act between data sources and the data warehouse [8]. Interestingly, the semantic annotation of the repositories may be aided by visual tools thus, intuitively and avoiding time-consuming efforts.

### 2.7. Shortcoming: Semantically-traceable Models

*Mapping from multidimensional models to relational models is accomplished by structural matching. Since structure and semantics are closely related in such a way, whenever structure is manually changed, the source semantics are lost.*

### 2.7.1. Description

The main different between model transformations and ontologies is that ontological mapping uses the explicitly codified knowledge whereas models have to assume the meaning from the codification of the modeling [33]. Multidimensional models are then concerned about data structures for OLAP purposes. The key idea is that model-to-model transformations have been designed for matching structures instead of matching semantics as ontologies do.

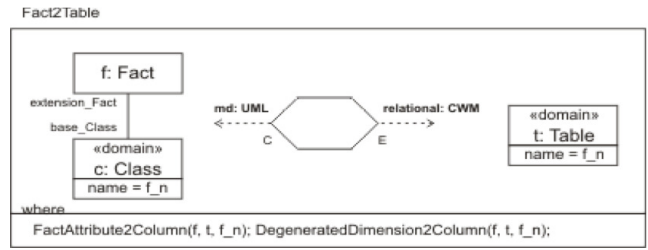

Figure 1. QVT relation for mapping facts into tables

Figure 2 shows the usual situation where multidimensional modeling elements (facts in the figure) are related to others (a table in a relational deployment of the data warehouse). Despite of the fact that Query/View/Transformation (QVT) language (as appear in [4]) codifies this relation, the structural matching is articulated using simple name matching/comparison (using string equality). It is illustrated by the *name=$f_n$* rule in Figure 2. A QVT engine can match here complex data structures taking into account these kinds of relations. However, because the matching is indeed structural, these labels acting as structural identifiers do not provide suitable mappings when they come from arbitrary decisions. Therefore, data warehouse designers may change them indiscriminately along time.

### 2.7.2. Discussion

Automatic model transformations are managed with the manual transformations that software engineers do, *i.e.*, modify the current label of a modeling element, remove some data, relate some elements together, and so on. Traceability is the mechanism for propagating changes in a transformation chain taking into account only the changes and not entire models (this capability is referred in programming as incremental compilation). It solves the problem whenever automatic transformations are involved. However, modifying the label implies change its semantics or even worse, losing its meaning because the misuse of semantics in model-driven methods. This fact is easily shown in data sources where a data item may be label in a cryptic form (*s0702*) on which its semantics cannot be inferred (*the sales from 2007 to 2009*).

Ontologies may solve this drawback in model-driven development by semantically annotating multidimensional models. Then, instead of labeling them for a dual purpose, *i.e.*, specify semantics and identifying structures, each modeling element keeps an (ontological) identity about what is their meaning. In Figure 2, identifiers in the sense of ontology references should be used instead of *name*. Then, structural mappings such as this will properly manage changes in structure (*e.g.*, caused by user manipulation) without affecting semantics. In this case, string equality is converted into identifier/reference equality that may be easily mapped to a numeric equality (solving the exact matching required by string equality or complex auxiliary similarity functions).

### 2.8. Shortcoming: Reasoning on OLAP Queries

*OLAP queries are based on the manipulation of data aggregations. However, OLAP algebras are based on calculus instead of logics. Any proposition about the multidimensional model cannot be proven but just calculated.*

### 2.8.1. Description

The main insight is to enrich OLAP-querying expressiveness with the inference capability provided by the ontological foundation of the data warehouse. Similarly to the geographical enrichment of traditional OLAP systems [34] or the alignment of data warehouses with document databases [35], the ontological query may provide interesting capabilities to OLAP analysis.

### 2.8.2. Discussion

Take an OLAP analysis over commercial transactions as example. These are related to some dimensions such as *product*, *customer*, and *season*. Despite of the fact that OLAP analysis enables decision makers to explore hypothesis concerning these dimensions (*e.g.*, such-and-so customer kinds determine such-and-so commercial transaction), inferring knowledge from the logical propositions provided by ontologies is still missing (*e.g.*, is a kind of customer logically associated to a product type regarding a given season?).

Moreover, semantics-aware querying of data warehouses provides decision makers with ontologically-founded models. Decision makers can properly interpret these models because they have been obtained by agreement.

### 2.9. Shortcoming: Asserting Suitable Visualizations

*Deciding which visualization may be suitable for a given data set is one of the top 10 problems in the visualization information field [36]. Data cubes can be visualized in this sense using many techniques. Finding ways of aiding decision makers to select the proper visualization will ameliorate this task.*

### 2.9.1. Description

What kind of visualization technique is provided? Regarding the very nature of data cubes is another topic on which the semantic data annotation may aid to improve the development process. On the one hand, data warehouses are multidimensionally modeled, *i.e.*, describing facts and dimensions of analysis. On the other hand, visualization techniques have particular information requirements in terms of the data types. For instance, Table 1 shows many of the visualizations used by Many Eyes software solution (http://manyeyes.alphaworks.ibm.com). This application presents a method for visualizing data sets by novels. Interestingly, taking Many Eyes as example, techniques related to the category *see the world* (see Table 1) requires data under the signature $\Sigma^* \times R$.

### 2.9.2. Discussion

The semantic annotation of measures and dimensions enables the automatic matching of these visualization information requirements with the multidimensional structure of the data warehouse. For instance, given the semantic category of a concept in a multidimensional model, its characterization as a categorical data (denoted as $\Sigma *$ in Table 1) or as a ratio scale (R) may be ontologically inferred.

Table 1. Information requirements for the visualization techniques used by Many Eyes. It is assumed that R is ordered and $\Sigma*$ has no order defined. *m*, *n* are non-negative natural numbers.

| Category | Name | Format |
|---|---|---|
| See the world | World map | $\Sigma* \times R$ |
| | Country maps | $\Sigma* \times R$ |
| Track rises and falls over time | Line graph | $R \times R^n$ |
| | Stack graph | $R \times R^n$ |
| | Stack graph for cat. | $(\Sigma*)^m \times R^n$ |
| Compare a set of values | Bar chart | $\Sigma* \times R^n$ |
| | Block histogram | $\Sigma* \times R$ |
| | Bubble chart | $\Sigma* \times R$ |
| | Matrix chart | $(\Sigma*)^j \times R^k: j > 1$ and $k \in \{0,1\}$ |
| See relationships among data pts. | Scatterplot | $\Sigma* \times R^3$ |
| | Network diagram | $(\Sigma*)^2$ |
| See the parts of a whole | Pie chart | $\Sigma* \times R^n$ |
| | Treemap | $(\Sigma*)^m \times R$ |
| | Change treemap | $(\Sigma*)^m \times R^2$ |
| Look for common words in a text | Tag cloud | $\Sigma* \times R$ |
| | Word tree | $\Sigma*$ |

## 2.10. Shortcoming: Security Constraint Validation

*Security is commonly specified in data warehouses in an* ad-hoc *basis. Security constraints can be inferred by the ontological reasoning about credentials, permissions and rights.*

### 2.10.1. Description

Security is recognized as one main concern on which a data warehouse should be aware. Data warehouses are large repositories of sensible data that may drive organizations to make important decisions. Therefore, security issues such as access control and audition are as critical as the actual data quality that the data warehouse and ETL processes can assure.

An archetypal example of what is followed up till now comes from works such as [37]. These authors define an access control and audit (ACA) model for data warehouses. This model enriches the multidimensional structures of an OLAP model with the information describing users, privileges and policies that finally serve to articulate an ACA policy. These policies are focused on the derivation of platform-specific security policies. For instance, if ORACLE is chosen as the target platform for implementing the data warehouse, the ACA model provided by [37] will be translated into the *ORACLE Label Security*.

International Journal of Database Management Systems ( IJDMS ), Vol.3, No.2, May 2011

**2.10.2. Discussion**

Models such as ACA may be generalized as ontologies. In fact, there are several ontologies already proposed for capturing the semantics involved. Some of them are [38] proposing a credentials ontology oriented to the domain of web services and requesting agents based on DAML (http://www.daml.org), the *eXtensible Access Control Markup Language* (XACML) or the NRL Security Ontology [39]. For instance, the last one, a comprehensive ontology composed by several security-related ontologies such as the main security ontology (describing security concepts), credentials ontology (specifying authentication credentials), etc.

Importantly, security ontologies open the door to the reasoning about security constraints and privileges. First, security constraints are specified without any aid but the actual knowledge of data modelers in model-driven approaches such as [37]. However, thanks to the ontological definitions of these constraints, the provided reasoning capabilities can check conflicts among constraints. Moreover, access policies may be articulated by reasoning. Data warehouse models may be marked with security annotations (that act as authorization rules <*target, effect, condition*>) that refer to the selected ontology and then, by matching these rules to the user credentials, the access control may be resolved. For this aim, security reasoners decide relationships between information requirements and user capabilities using automatic subsumption. For instance, security reasoning on multidimensional models is managed along aggregation hierarchies: if a given set of privileges do not permit that a user access to a particular aggregation level data, this restriction should be properly *propagated* to the rest of the hierarchy (to upper levels for allowing access and to the lower for denying it). In the propagation of authorization rules ontologies thus may formalize the *ad-hoc* mechanisms encountered in the current literature.

## 3. CONCLUSIONS

This article is focused on describing how data warehouse design can take advantage of ontologies. We have described several shortcomings that we have detected during the development of several real-world projects. Furthermore, we do not only describe each shortcoming but also we discuss how ontologies can be considered to overcome them. Ontologies formalize specific-domain knowledge that may benefit data warehousing by:

- reusing expert knowledge from different domains,

- enriching specific metadata by completing definitions and annotating their semantics,

- enabling metadata interchange among repositories,

- populating the designed databases from public data sources,

- empowering data integration, and analysis,

- automatizing reasoning on metadata,

- validating data instances and models, and

- helping to understand the meaning of the notions that are actually modeled.

84

Therefore, we have given reasons for the convenience of managing ontological knowledge and tools for ameliorating many of the heavy tasks involved in data warehousing. This article pretends to start such discussion, thus posing a starting point for further research in the area, by also considering other related aspects, such as schema evolution [40], data quality [41], or data fusion [42].

## ACKNOWLEDGEMENTS

Spanish Ministry of Education and Science funds Jesús Pardillo under the FPU grant AP2006-00332. Special thanks to Juan Trujillo for his comments on the draft of this article.

## Authors

**Jesús Pardillo** graduated with honours in Information Systems Engineering in 2006 and obtained his Ph.D. in Computer Science in 2010 from the University of Alicante. He has publications on software engineering in international conferences such as MODELS, DOLAP, QUATIC, ER, DaWaK, or JISBD, and journals such as the Journal of Systems and Software, the International Journal of Intelligent Systems, Information Sciences, or Information and Software Technology. His current research interests include: foundations of mathematics, knowledge management, programming languages, and information visualization. Contact him at research@jesuspardillo.com or follow him at www.jesuspardillo.com.

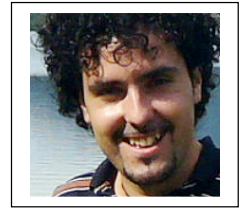

**Jose-Norberto Mazón** is Assistant Professor at the Department of Software and Computing Systems in the University of Alicante (Spain). He obtained his Ph.D. in Computer Science from the University of Alicante (Spain) within the Lucentia Research Group. He has published several articles about data warehouses and requirement engineering in national and international workshops and conferences, (such as DAWAK, ER, DOLAP, BNCOD, JISBD and so on) and in several journals such as Decision Support Systems (DSS), SIGMOD Record or Data and Knowledge Engineering (DKE). He has also been co-organizer of the International Workshop on Business intelligencE and the WEB (BEWEB 2010) and the International Workshop on The Web and Requirements Engineering (WeRE 2010). His research interests are: business intelligence, design of data warehouses, multidimensional databases, requirement engineering and model driven development. Contact him at jnmazon@dlsi.ua.es.

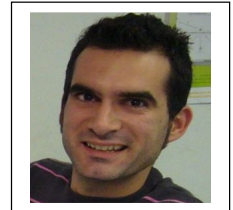